\documentclass[journal]{IEEEtran}

\ifCLASSINFOpdf
\else
   \usepackage[dvips]{graphicx}
\fi
\usepackage{url} 

\usepackage[utf8]{inputenc}
\usepackage{graphicx}
\usepackage{amsmath}
\usepackage{float}
\usepackage{array, multirow, booktabs}
\usepackage{amssymb}
\usepackage[ruled,vlined]{algorithm2e}
\usepackage{placeins}
\usepackage[hidelinks]{hyperref}
\usepackage{footmisc}

\DeclareMathOperator*{\argmin}{arg\,min}

\begin{document}

\title{Adversarial Unsupervised Domain Adaptation for Harmonic-Percussive Source Separation}

\author{Carlos Lordelo$^{1,2}$, Emmanouil Benetos$^1$, Simon Dixon$^1$, Sven Ahlbäck$^2$, and Patrik Ohlsson$^2$\\
\vspace{0.15cm}$^1$ Centre for Digital Music, Queen Mary University of London, UK\\
$^2$ Doremir Music Research AB, Stockholm, Sweden
\thanks{This work has received funding from the European Union's Horizon 2020 research and innovation programme under the Marie Skłodowska-Curie grant agreement No.~765068 - \href{https://mip-frontiers.eu/}{MIP Frontiers Project}.}}

\maketitle

\begin{abstract}
This paper addresses the problem of domain adaptation for the task of music source separation. Using datasets from two different domains, we compare the performance of a deep learning-based harmonic-percussive source separation model under different training scenarios, including supervised joint training using data from both domains and pre-training in one domain with fine-tuning in another. We propose an adversarial unsupervised domain adaptation approach suitable for the case where no labelled data (ground-truth source signals) from a target domain is available. By leveraging unlabelled data (only mixtures) from this domain, experiments show that our framework can improve separation performance on the new domain without losing any considerable performance on the original domain. The paper also introduces the Tap \& Fiddle dataset, a dataset containing recordings of Scandinavian fiddle tunes along with isolated tracks for ``foot-tapping" and ``violin".
\end{abstract}


\IEEEpeerreviewmaketitle
\vspace{-0.5cm}
\section{Introduction}
\label{sec:introduction}
\IEEEPARstart{B}{lind} source separation (BSS) is a fundamental problem in signal processing. It consists of separating a set of mixture signals into a set of source signals without using any extra information 
\cite{Comon10}. In this work, we will be considering the task of Music Source Separation (MSS), which is an ill-posed and underdetermined case of BSS, where multiple sources (instrumental signals) must be separated from a single mixture (music recording).
Current MSS methods are based on Deep Neural Networks (DNNs) that need a lot of labelled data (mixtures and ground-truth isolated instrumental signals) to be trained under a supervised scenario \cite{openUnmix, Takashi18}. However, labelled audio data for MSS is difficult to obtain.
In the literature, there are only a few large-scale public datasets for MSS, such as MUSDB18 \cite{musdb18} and Slakh \cite{Manilow2019}. 

Even though it is known that the use of data augmentation techniques such as random pitch-shifting and random mixing of source signals can improve model generalisation \cite{Uhlich2017, Cohen-Hadria2019}, separation performance will always depend on the type of audio data used during training. When the data distribution of the training set is different from the data distribution of the test set, the performance of any predictor is degraded. This effect is known as dataset shift \cite{Quionero-Candela2008}, and happens due to mismatched characteristics between data used for training and testing.

Under this scenario, domain adaptation techniques address this problem by adapting predictors from a \textit{source domain}, where usually a large amount of labelled data is available, to a \textit{target domain}, where only few or no labelled data is available. Domain adaptation is already consolidated as an important research topic in computer vision, where it is used in complex classification tasks
\cite{Peng2019}. Even in closer fields, such as acoustic scene analysis \cite{Gharib2018, Wei2020}, speech recognition \cite{Sun2017} and speech enhancement \cite{Meng2018}, 
domain adaptation methods have already been proposed. However, to our knowledge, methods of this nature have not yet been investigated for MSS. Therefore, our work also attempts to fill this gap in the literature.

We propose an adversarial unsupervised domain adaptation approach for MSS. By using the mixtures and the available ground-truth signals from MUSDB18 and a set of unlabelled data (mixtures) from a different domain, we show that our framework is able to improve separation performance in the new domain while maintaining the original performance on MUSDB18, considerably reducing the degradation effect caused by dataset shift. Although our experiments are carried out for the particular task of Harmonic-Percussive Source Separation (HPSS), our framework can be easily adapted to other MSS tasks with different types of sources and domains. 

In summary, our contributions include:\begin{itemize}
    \item The first work focused on unsupervised domain adaptation for MSS;
    \item An adversarial unsupervised domain adaptation framework for MSS that can be used with any neural network architecture, any type of audio representation and any number of sources;
    \item The public release of the ``Tap \& Fiddle Dataset", a dataset containing recordings of traditional Scandinavian fiddle tunes with accompanying foot-tapping along with isolated tracks for ``foot-tapping" and ``violin".
    This dataset has different timbral characteristics than MUSDB18 and is useful for domain adaptation experiments;
    \item A prototype experiment where we show an improvement over benchmark methods for the HPSS task.
\end{itemize}

\vspace{-0.1cm}
\section{Related Work}
\label{sec:bib_review}
\subsection{Harmonic-Percussive Source Separation}

The task of HPSS consists of separating a music signal into two source signals, one with the harmonic components and other with the percussive sounds \cite{Ono2008}. Signal processing methods for HPSS perform separation by exploiting the fact that percussive signals form vertical lines in the mixture time-frequency representation, while the harmonic components tend to form horizontal structures, e.g. \cite{Fitzgerald10, Driedger14_HPSS, Liutkus14_KAM}. 
However, due to their strict assumptions and hand-crafted features, methods of this nature have intrinsic performance limitations.

Over the years, data-driven approaches have shown significant improvements over traditional methods for HPSS and current state-of-the-art methods are based on DNNs \cite{Roma2018, Lim17, Drossos18, Lordelo19}. In previous work carried out by the authors \cite{Lordelo19}, the $3$W-MDenseNet, an encoder-decoder DNN that uses convolutions with several kernel shapes to perform HPSS, was proposed. In this work, the same architecture is used, but here we add a domain discriminator into the framework and modify the loss function to support adversarial domain adaptation.

Moreover, since our approach is also grounded in Generative Adversarial Networks (GANs) \cite{Goodfellow2014}, it is important to point out some key aspects in which our proposal is different from other GAN-based source separation methods \cite{Stoller2018, Fan2018, Ong2019}.
\subsubsection{Discriminator}{Works on GAN-based MSS use a \emph{source discriminator}, which is trained to differentiate \textit{real} source signals from \textit{fake} source signals. This is different from our work, where we use a \emph{domain discriminator} trained to differentiate mixtures across two different domains.}
\subsubsection{Unlabelled data}{In order to train a source discriminator, a large number of single-source signals are required, even though those signals do not necessarily have to be paired with a music mixture. Here, we only need mixtures from each of the two domains to successfully train our domain discriminator.} \subsubsection{Input to discriminator}{The input to a source discriminator of GAN-based MSS works is the \emph{output} of the separator network. Our approach applies the domain discriminator on the \emph{encoded feature-maps}, in the middle of the separator network and not directly on its output.}

\vspace{-0.15cm}\subsection{Domain Adaptation}
\label{subsec:domain_adap}
Domain adaptation methods can be either supervised or unsupervised depending on the type of data from the target domain that is used. While Supervised Domain Adaptation (SDA) methods use labelled data, Unsupervised Domain Adaptation (UDA) exploits only unlabelled data (mixture signals) from the target domain. 

A typical SDA approach is to first train a model using a large number of labelled samples from the source domain and then re-train some (or all) of its layers using a smaller labelled dataset of interest (target domain). This technique is known as \emph{fine-tuning} 
\cite{Oquab2014, Wang2018}. Another SDA approach is \emph{joint training}, where the two datasets are merged into a new dataset and only a single training stage is done, using labelled data from both domains in every batch \cite{Maciejewski2018, Manilow2019}.

UDA methods usually consider that the system is under the \emph{covariate shift paradigm}, assuming that, even though the marginal distribution of source domain data is different from the marginal distribution of target domain data, the conditional probability of the output remains the same. 
Therefore, if the marginal distributions can be matched, the same predictor can be applied successfully over samples from either of the two domains \cite{Shimodaira2000}. 
In order to do this, some UDA methods propose to re-weight \cite{Huang2006} or select samples from the source domain \cite{Gong2013}, while others project the data through an embedding function such that not only the marginals become similar on the embedded space, but also the embedded features keep their discrimination potential \cite{Fernando2013, Xiao2018}. 
The latter case is also the type of UDA method in our proposal. We look for a transformation that creates an embedded space in which the confusion between the two domains is maximised.

Similar to \cite{Ganin2015}, we propose to find a \emph{domain-invariant} and \emph{separation-discriminative} embedded space that is learned from data via adversarial training. 
However, differently from \cite{Ganin2015}, we deal with the task of source separation (regression) instead of image recognition (classification). 
In addition, we use CNNs for the encoder-decoder and the domain discriminator, while in \cite{Ganin2015} simple feed-forward networks are used, and while \cite{Ganin2015} performs adversarial training using the gradient reversal layer method, we conduct conditional GAN iterative optimisation as in \cite{Goodfellow2014}. 

\section{Proposed Framework}
\label{sec:Proposed Model}
We assume that both the input data and the outputs are $F\!\times\!T$ magnitude spectrograms, where $F$ is the number of frequency bins and $T$ the number of frames. To simplify the notation, we treat them as vectors in $\mathbb{R}^{K}$, where $K = FT$. Hence, the input (mixture signal) is notated as $\mathbf{x}$ and its labels (ground-truth isolated source signals) as the $K\!\times\!2$ matrix $\mathbf{Y} = \big[\mathbf{h} ~ \mathbf{p}\big]$, where the first column is the original harmonic vector $\mathbf{h} \in \mathbb{R}^K$ and the second column is the original percussive vector $\mathbf{p} \in \mathbb{R}^K$. Furthermore, we consider that the mixture-label pairs follow the joint distribution $p_\mathcal{A}(\mathbf{x}, \mathbf{Y})$, or, in other words, we say that the data ``come from  domain $\mathcal{A}$". For the general supervised HPSS case, the goal is to train a model based on this data that can be a good predictor of $p(\mathbf{Y} | \mathbf{x} \sim p_\mathcal{A}(\mathbf{x}))$. 

In \cite{Lordelo19} we proposed the $3$W-MDenseNet, a convolutional encoder-decoder for HPSS, where the network output is an estimate $\hat{\mathbf{Y}} = \big[\hat{\mathbf{h}} ~ \hat{\mathbf{p}}\big]$ of $\mathbf{Y}$.
Here, we model the encoder-decoder-based separation process as a sequence of two mappings. First, the encoder $\mathcal{E}$ with parameters $\theta_{\mathcal{E}}$ maps the input to an embedded feature space $\mathbf{z} = \mathcal{E}(\mathbf{x}; \theta_{\mathcal{E}})$ and then the decoder $\mathcal{D}$, with parameters $\theta_{\mathcal{D}}$, maps $\mathbf{z}$ to the output $\hat{\mathbf{Y}}$ such that: 
\begin{equation}
    \hat{\mathbf{Y}} = \mathcal{D}(\mathbf{z}; \theta_{\mathcal{D}}) = \mathcal{D}(\mathcal{E}(\mathbf{x}; \theta_{\mathcal{E}}); \theta_{\mathcal{D}}).
\end{equation} This separator can be optimised for the general supervised HPSS case using the mean square error as the loss $\mathcal{L}\strut_{\mathrm{S}}$ \cite{Lordelo19}:
\begin{multline}
    \!\!\mathcal{L}\strut_{\mathrm{S}}(\theta_{\mathcal{E}}, \theta_{\mathcal{D}}) = \hspace{-0.7cm}\underset{~{}~{}~{}~{}\mathbf{x} \sim p_{\mathcal{A}}(\mathbf{x})}{\mathbb{E}} \hspace{-0.6cm}\big[\lambda_{h}||\hat{\mathbf{h}} - \mathbf{h}||^2 + \lambda_{p}||\hat{\mathbf{p}} - \mathbf{p}||^2\big] =\hspace{-0.6cm}\underset{~{}~{}~{}~{}\mathbf{x} \sim p_{\mathcal{A}}(\mathbf{x})}{\mathbb{E}}\hspace{-0.6cm}\big[|| (\hat{\mathbf{Y}} \\ - \mathbf{Y})\mathbf{\Lambda}||_{F}^2\big] = \!\hspace{-0.6cm}\underset{~{}~{}~{}~{}\mathbf{x} \sim p_{\mathcal{A}}(\mathbf{x})}{\mathbb{E}}\hspace{-0.6cm}\big[\!\left|\left|(\mathcal{D}(\mathcal{E}(\mathbf{x}; \theta_{\mathcal{E}}); \theta_{\mathcal{D}})\! -\! \mathbf{Y})\mathbf{\Lambda}\right|\right|_{F}^2\!\big], \label{eq:loss_mse}
\end{multline}
where $\lambda_{h}$ and $\lambda_{p}$ are weights for the harmonic and percussive outputs respectively --- we use $0.5$ for each since we want to assign equal importance to each source --- , $||\dots||$ represents the Euclidean norm, $||\dots||_{F}$ the Frobenius norm and $\mathbf{\Lambda}$ is the diagonal matrix $\big[\begin{smallmatrix}
  \sqrt{\lambda_{h}} & 0\\
  0 & \sqrt{\lambda_{p}}
\end{smallmatrix}\big]$.

However, in this work we assume there also exists a new domain $\mathcal{B}$, where mixtures follow the marginal distribution $p_\mathcal{B}(\mathbf{x})$, which is considered different from $p_\mathcal{A}(\mathbf{x})$. Our main goal is now to be able to robustly predict labels $\hat{\mathbf{Y}}$ given that the input can be from either domain $\mathcal{A}$ or $\mathcal{B}$. Apart from the labelled samples from domain $\mathcal{A}$, we have access to set of mixtures from $\mathcal{B}$ that can be used for performing UDA.

\begin{figure}[t]
  \centering
  \includegraphics[width=0.9\columnwidth]{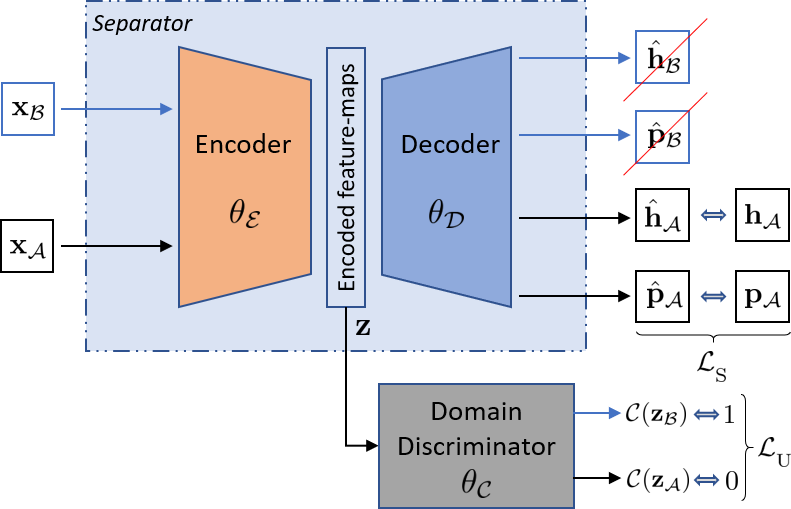}
  \caption{Schematic of proposed adversarial UDA for HPSS.}
  \label{fig:HPSS_domain_adapt}
\end{figure}

Our approach adopts a similar methodology to \cite{Ganin2015} and \cite{Ganin2016}. We propose to learn encoded features $\mathbf{z}$ that can not only guarantee a good separation performance, but that are also invariant to domain changes. This means that $\mathbf{z}$ must not contain any discriminative information about the origin of the input ($\mathcal{A}$ or $\mathcal{B}$). By doing so, we can make the distributions $p(\mathbf{z}|\mathbf{x}\!\!\sim\!\!p_{\mathcal{A}}(\mathbf{x}))=\{\mathcal{E}(\mathbf{x};\theta_{\mathcal{E}})|\mathbf{x}\!\sim\!p_{\mathcal{A}}(\mathbf{x})\}$ and $p(\mathbf{z}|\mathbf{x}\!\!\sim\!\!p_{\mathcal{B}}(\mathbf{x})) = \{\mathcal{E}(\mathbf{x}; \theta_{\mathcal{E}})|\mathbf{x}\!\sim\!p_{\mathcal{B}}(\mathbf{x})\}$ to become as similar as possible. In order to measure their similarity, we use a domain discriminator $\mathcal{C}(\mathbf{z}, \theta_{\mathcal{C}})$ to discriminate the encoded feature-maps between the two domains.
Such domain discriminator is a binary classifier that can be trained using only mixture signals by minimising the binary cross-entropy $\mathcal{L}\strut_{\mathrm{U}}$: 
\begin{equation}
     \mathcal{L}\strut_{\mathrm{U}}\!(\theta_{\mathcal{C}}, \theta_{\mathcal{E}})\!=\! 
    -\hspace{-0.6cm}\underset{~{}~{}~{}~{}\mathbf{z} \sim p_{\mathcal{B}}(\mathbf{z})}{\mathbb{E}} \hspace{-0.6cm}\big[\!\log \mathcal{C}(\mathbf{z}, \theta_{\mathcal{C}})\big]
     -\hspace{-0.66cm}\underset{~{}~{}~{}~{}\mathbf{z} \sim p_{\mathcal{A}}(\mathbf{z})}{\mathbb{E}}\hspace{-0.6cm}\big[\!\log(1 - \mathcal{C}(\mathbf{z}, \theta_{\mathcal{C}}))\big]\!.
     \label{eq:loss_disc}
\end{equation}
Fig.~\ref{fig:HPSS_domain_adapt} summarises the domain adaptation scenario.

In addition, we ensure that $\mathbf{z}$ will become domain-invariant by forcing the encoder sub-network to generate feature-maps that can fool the domain discriminator. This is achieved by maximising $\mathcal{L}\strut_{\mathrm{U}}$ when training the encoder weights. Such a min-max game is played by the encoder sub-network and the domain discriminator during training just like in GAN training \cite{Goodfellow2014}. 
At the same time, $\mathbf{z}$ can keep its separation-discriminative properties if we include the minimisation of $\mathcal{L}\strut_{\mathrm{S}}$ in the loss function. The final encoder loss is, therefore, a combination of the (unsupervised) \emph{adversarial} loss $\mathcal{L}\strut_{\mathrm{U}}$, which can be optimised using only mixture signals from each of the two domains, and the (supervised) loss $\mathcal{L}\strut_{\mathrm{S}}$, which can be optimised based only on samples from $\mathcal{A}$ since it requires labelled data. In summary, the loss functions of each sub-network are:
\begin{align}
     \hat{\theta}_{\mathcal{C}} &= \argmin_{\theta_{\mathcal{C}}}\mathcal{L}\strut_{\mathrm{U}}(\theta_{\mathcal{E}}, \theta_{\mathcal{C}}) \\
     \hat{\theta}_{\mathcal{E}} &= \argmin_{\theta_{\mathcal{E}}}\big[ 
     - \gamma_{\mathrm{U}}\mathcal{L}\strut_{\mathrm{U}}(\theta_{\mathcal{E}}, \hat{\theta}_{\mathcal{C}}) +
     \gamma_{\mathrm{S}}\mathcal{L}\strut_{\mathrm{S}}(\theta_{\mathcal{E}}, \hat{\theta}_{\mathcal{D}}) \big] \\
     \hat{\theta}_{\mathcal{D}} &= \argmin_{\theta_{\mathcal{D}}}\mathcal{L}\strut_{\mathrm{S}}(\theta_{\mathcal{E}}, \theta_{\mathcal{D}})
     \label{eq:loss_total}
\end{align} where $\gamma_{\mathrm{U}}$ and $\gamma_{\mathrm{S}}$ are weights given to the unsupervised part and to the supervised part of the loss. 

It should be noted that $\mathcal{C}$, $\mathcal{E}$ and $\mathcal{D}$ must be trained together in an iterative way as in GAN training \cite{Goodfellow2014}. If $\mathcal{C}$ is optimised to completion, the encoder sub-network will not be able to increase the domain-discriminator confusion, causing the separator performance to overfit over domain $\mathcal{A}$ \cite{Goodfellow2014}. In our experiments, at every training iteration, we perform $5$ updates on $\theta_{\mathcal{C}}$ before updating $\theta_{\mathcal{E}}$ and $\theta_{\mathcal{D}}$. The full training algorithm can be found in the supplementary material of this paper.

\begin{table*}[htb]
\centering
\caption{Objective evaluation of HPSS on MUSDB18 and Tap \& Fiddle. The values are in dB and represent the median of metrics over tracks in each test set. IBM is the Ideal Binary Masking and IRM represents the Ideal Ratio Masking oracle methods.}
\resizebox{0.978\textwidth}{!}{%
\begin{tabular}{ccccccccccccccc}
\toprule
\multicolumn{1}{c}{\multirow{4}{*}{Method}} & \multicolumn{12}{c}{Test Set} & \multicolumn{2}{c}{\multirow{2}{*}{Type of}}\\ \cmidrule{2-13}
  & 
\multicolumn{6}{c|}{MUSDB18 (Domain $\mathcal{A}$)} & \multicolumn{6}{c}{Tap \& Fiddle (Domain $\mathcal{B}$)} & \multicolumn{2}{c}{Data} \\  
(Training Set) & \multicolumn{3}{c}{Percussive} & \multicolumn{3}{c|}{Harmonic} & \multicolumn{3}{c}{Percussive} & \multicolumn{3}{c}{Harmonic} & \\ \cmidrule{2-15} 
& SDR & SIR & SAR & SDR & SIR & SAR & SDR & SIR & SAR & SDR & SIR & SAR & $\mathcal{A}$ & $\mathcal{B}$ \\ \toprule
HPSS\_MUSDB  ($\mathcal{A}$) & $4.5$ & $13.0$ & $5.0$ & $10.0$ & $13.4$ & $12.3$ & $1.3$ & $15.8$ & $0.3$ & $22.0$ & $23.0$ & $29.7$ & labelled & ---  \\ \midrule
HPSS\_T\&F ($\mathcal{B}$) & $-0.2$ & $0.3$ & $10.5$ & $3.1$ & $16.5$ & $5.2$ & $10.2$ & $16.9$ & $12.7$ & $35.0$ & $36.4$ & $34.3$ & --- & labelled\\ \midrule
SDA\_joint ($\mathcal{A} + \mathcal{B}$) & $4.8$ & $13.3$ & $5.1$ & $10.2$ & $13.9$ & $12.1$ & $4.6$ & $18.1$ & $6.4$ & $27.5$ & $28.9$ & $30.2$ & labelled & labelled  \\ \midrule
SDA\_tuned ($\mathcal{A} \rightarrow \mathcal{B}$) & $2.9$ & $8.6$ & $3.3$ & $7.1$ & $9.3$ & $10.5$ & $12.1$ & $18.8$ & $12.6$ & $35.3$ & $37.1$ & $35.6$ & labelled & labelled\\ \toprule
\textbf{UDA\_small}  & $4.8$ & $12.2$ & $5.1$ & $10.0$ & $13.4$ & $11.8$ & $3.4$ & $13.0$ & $2.9$ & $25.0$ & $25.9$ & $30.8$  & labelled & unlabelled  \\ \midrule
\textbf{UDA\_large}  & $4.6$ & $12.9$ & $4.9$ & $10.1$ & $14.1$ & $12.0$ & $7.4$ & $18.0$ & $8.4$ & $29.2$ & $30.6$ & $33.1$ & labelled & unlabelled\\ \toprule
OpenUnmix \cite{openUnmix} & $5.2$ & $11.2$ & $6.0$  & $10.1$ & $17.7$ & $10.7$ & $6.7$ & $7.0$ & $5.1$ & $28.6$ & $36.8$ & $25.9$ & labelled & ---\\ \midrule
IBM  & $7.8$ & $16.4$ & $7.9$ & $11.9$ & $17.9$ & $13.2$ &  $13.5$ & $20.8$ & $13.7$ & $37.8$ & $41.3$ & $37.7$ & --- & ---  \\ \midrule
IRM  & $8.0$ & $12.4$ & $9.7$ & $12.2$ & $15.8$ & $15.0$ & $13.4$ & $19.5$ & $13.8$ & $37.2$ & $42.0$ & $37.2$ & --- & --- \\ \bottomrule
\end{tabular}%
}
\label{tab:results}
\end{table*}

\section{Datasets}
MUSDB18 \cite{musdb18} is the largest public dataset for MSS containing real-world audio recordings. It contains full-track songs and includes both the mixtures and the original sources, divided between a training subset of $100$ music recordings and a test subset of $50$.
The available isolated tracks are vocals, bass, drums and ``other". We use the drum track as the ground-truth for the percussive source, while the sum of the other tracks is used as ground-truth for the harmonic source.

As a different domain, we collected and publicly release the Tap \& Fiddle (T\&F) dataset \cite{LordeloTF20}. The T\&F dataset contains stereo recordings of traditional Scandinavian fiddle tunes with accompanying foot-tapping, which is standard performance practice within these musical styles. It consists of $28$ recordings with completely separate fiddle and foot-tapping sounds as well as mixed signals. The dataset is divided into a training set with $23$ files and a test set with $5$. All recordings are solo and have an average duration of $65$ seconds.
Detailed information regarding the T\&F Dataset can be found in \cite{LordeloTF20}. 

\section{Experimental Setup}
In our experiments, the music signals are converted to mono and resampled to $16$KHz. The inputs are normalised magnitude spectrograms of size $256\times256$ generated by the application of an STFT of size $512$ with $75$\% overlap. A validation split of $20\%$ of all labelled data available for training is set.

\begin{figure}[htb]
  \centering
  \includegraphics[width=\columnwidth]{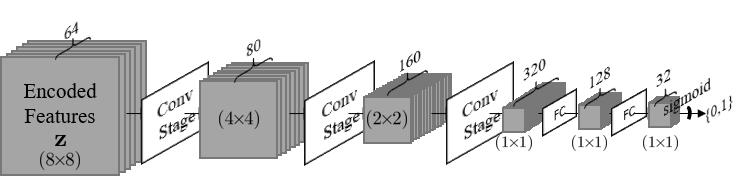}
  \caption{Architecture of the domain discriminator. Each ``Conv Stage" is a $3\times3$ convolutional layer followed by $2\times2$ max pooling. ``FC" is a fully connected layer.}
  \label{fig:disc_arch}
\end{figure}

We use the $3$W-MDenseNet \cite{Lordelo19} as the separator architecture. As a post-processing step, we apply Wiener filtering \cite{Nugraha2016} to the source estimates and use the mixture phase to return to the time domain. We concatenate the encoded feature-maps of each of the three branches of the $3$W-MDenseNet to form $\mathbf{z}$. Details about hyper-parameter choices can be found in the paper's supplementary material. The architecture of the domain-discriminator network is depicted in Fig.~\ref{fig:disc_arch}.  

After experimentation, we choose the values of $1$ for $\gamma_{\mathrm{S}}$ and $0.001$ for $\gamma_{\mathrm{U}}$. Training is performed using the Adam optimiser with an initial learning rate of $0.001$, which is reduced by a factor of $0.25$ if the supervised validation loss $\mathcal{L}\strut_\mathrm{S}$ stops improving for $50$ consecutive epochs, and if no improvement happens in $200$ epochs the training is stopped. The separation quality is evaluated using the BSS\_eval \cite{Vincent2006} set of objective metrics that are largely used by the MSS community. 

\section{Results}
\label{sec:results}
Recordings from MUSDB18 represent domain $\mathcal{A}$ while recordings from the T\&F dataset represent domain $\mathcal{B}$. We aim to investigate how different training scenarios perform across the two domains. We compare our UDA proposal to traditional supervised HPSS approaches that use only labelled data from one of the domains, to SDA frameworks, which include joint training using labelled data from both datasets and fine-tuning over samples from T\&F after training on MUSDB18, and to another state-of-the-art DNN for MSS named OpenUnmix \cite{openUnmix}. This method was previously trained on an augmented version of MUSDB18 and serves as a baseline in our comparison. 

In addition to the mixtures in the T\&F dataset, we have a collection of $50$ new recordings of Scandinavian fiddle tunes with accompanying foot-tapping. This collection is also part of domain $\mathcal{B}$ and although no labels are available, it can also be used by our UDA method. We then test two versions of our approach: HPSS\_UDA\_small, which uses the mixtures on the train set of T\&F for performing the adaptation to domain $\mathcal{B}$, and HPSS\_UDA\_large, which uses the larger set of mixtures from our internal collection. Results are shown in Table \ref{tab:results}. 

By inspecting Table \ref{tab:results}, we can readily note that models that were trained only with samples from one dataset had poor performance on the other, which makes it possible to conclude that MUSDB18 and T\&F have very different priors over the data. This fact is also reflected in the performance of OpenUnmix, which is much lower on T\&F if compared with the performance provided by the ideal masking methods. Moreover, as expected, the joint trained model, SDA\_joint, achieved relatively good performance overall because it uses supervised data from both domains. The SDA\_tune model, which is the HPSS\_MUSDB model fine-tuned for T\&F, was indeed greatly improved when evaluated over this domain, but, as a trade-off, it lost a lot of its original performance on the original MUSDB18 dataset. On the other hand, both versions of the proposed UDA approach got a boost in performance on all $3$ of the metrics on T\&F without losing any considerable performance on MUSDB18. This means that our proposed UDA approach can perform HPSS on both domains successfully, even though the labelled data used for training came only from domain $\mathcal{A}$. 

The quantity of unlabelled data from domain $\mathcal{B}$ also impacted the performance of the proposed method. Even though the results of UDA\_large are similar to UDA\_small over domain $\mathcal{A}$, the former performs much better over samples from domain $\mathcal{B}$ than the latter due to the fact that it uses more than double the amount of mixtures from this particular domain during training to perform domain adaptation. Another interesting result is that UDA\_large, which is a semi-supervised framework, had similar performance over MUSDB18, but much better over T\&F if compared to SDA\_joint, which is a fully supervised method. This means that UDA using large amounts of unlabelled data can be much more promising than joint training using a smaller amount of labelled data.

More information about our work can be found in the paper's supplementary document and supplementary webpage\footnote{\href{http://c4dm.eecs.qmul.ac.uk/auda-hpss}{http://c4dm.eecs.qmul.ac.uk/auda-hpss}}.

\section{Conclusions}
In this work we presented an adversarial UDA model for HPSS. Our proposal is a semi-supervised framework that is able to exploit unlabelled mixtures from a target domain in order to improve HPSS generalisation to samples from this particular domain.
Results showed that our framework improves separation performance on the target domain without losing considerable performance on  the source domain.

As future work, we plan to investigate how the utilisation of small amounts of labelled samples from the target domain affect domain adaptation performance. We believe that this ``few-shot" approach can be useful in improving source separation performance in the absence of many data samples.  

\bibliographystyle{IEEEtran}
\bibliography{refs}

\end{document}